\documentclass[11pt,twocolumn]{article}
\usepackage{fullpage}
\usepackage{amsmath}
\usepackage[margin=1.65cm]{geometry}
\usepackage{graphicx}
\title{\center\textbf{Metamaterial Van Hove Singularity}}
\author{\textbf{C. L. Cortes, Z. Jacob*} \\ 
Department of Electrical and Computer Engineering, University of Alberta,\\
Edmonton, AB T6G 2V4, Canada \\
\\
*zjacob@ualberta.ca}
\date{}
\begin{document}

\twocolumn[
  \begin{@twocolumnfalse}
    \maketitle
	\begin{abstract}

	We introduce the photonic analogue of electronic Van Hove singularities (VHS) in artificial media (metamaterials) with hyperbolic dispersion. Unlike photonic and electronic crystals, the VHS in metamaterials is unrelated to the underlying periodicity and occurs due to slow light modes in the structure. We show that the VHS characteristics are manifested in the near-field local density of optical states inspite of the losses, dispersion and finite unit cell size of the hyperbolic metamaterial. Finally we show that this work should lead to quantum, thermal, nano-lasing and biosensing applications of van hove singularities in hyperbolic metamaterials achievable by current fabrication technology.

	\end{abstract}
	\vspace{1cm}
  \end{@twocolumnfalse}
]











\section{Introduction}

Periodic electronic and photonic crystals support a host of phenomena associated with the bloch waves of the underlying lattice structure. Along with band-gaps which completely forbid propagating waves, there exist critical points where the band structure (dispersion relation of propagating waves) has an extremum. These extrema lead to Van Hove singularities (VHS) in the density of states of the medium since it is related to the band structure $E(k)$ by $\rho(E)=\int{\frac{dS}{|\nabla E(k)|}}$, where dS is iso-frequency surface element.  VHS in the electronic density of states significantly affects the optical absorption spectra of solids\cite{kataura1999optical}. 

In photonic crystals, this extremum in the band structure corresponds to slow light modes of the structure where the group velocity decreases significantly below c. In the ideal limit, slow light modes have a zero group velocity leading to Van Hove singularities (VHS) in the photonic density of states (PDOS), the physical quantity govering light matter interaction. Thus slow light can enhance light-matter interactions that are critical for many applications in non-linear optics\cite{krauss_slow_2007}, and quantum optics  \cite{baba2008slow}. Demonstrations of slow light and associated effects have grown rapidly over the past decade -- evolving from low-temperature experiments to room-temperature set-ups that utilize electromagnetically induced transparency (EIT), coherent population oscillations and photonic crystals (PhCs)\cite{boyd2011material, krauss_slow_2007,khurgin_slow_2010}.



Here, we show that metamaterials can also support VHS, however their origin is unrelated to the underlying subwavelength periodicity of the lattice and arises due to the effective medium response. The  VHS in metamaterial waveguide structures is due to a balance of energy flow both inside and outside the metamaterial leading to slow light modes\cite{hess_active_2012,alekseyev_slow_2006}. We consider a practical waveguide structure where the proposed enhancement in the density of states can be experimentally verified at optical frequencies. We characterize light-matter enhancement using the local density of states (LDOS) in the near-field which demonstrate large enhancements at slow-light mode wavelengths\cite{yao_ultrahigh_2009} inspite of material absorption, dispersion, and finite patterning scale.  We emphasize that our paper presents practical designs for the metamaterial waveguide and also consider applications in distinct areas of quantum optics, thermal engineering, nano-lasing and biosensing.

\begin{figure*}[t]

\begin{minipage}[b]{0.25\linewidth}
\begin{center}
\includegraphics[width=47mm]{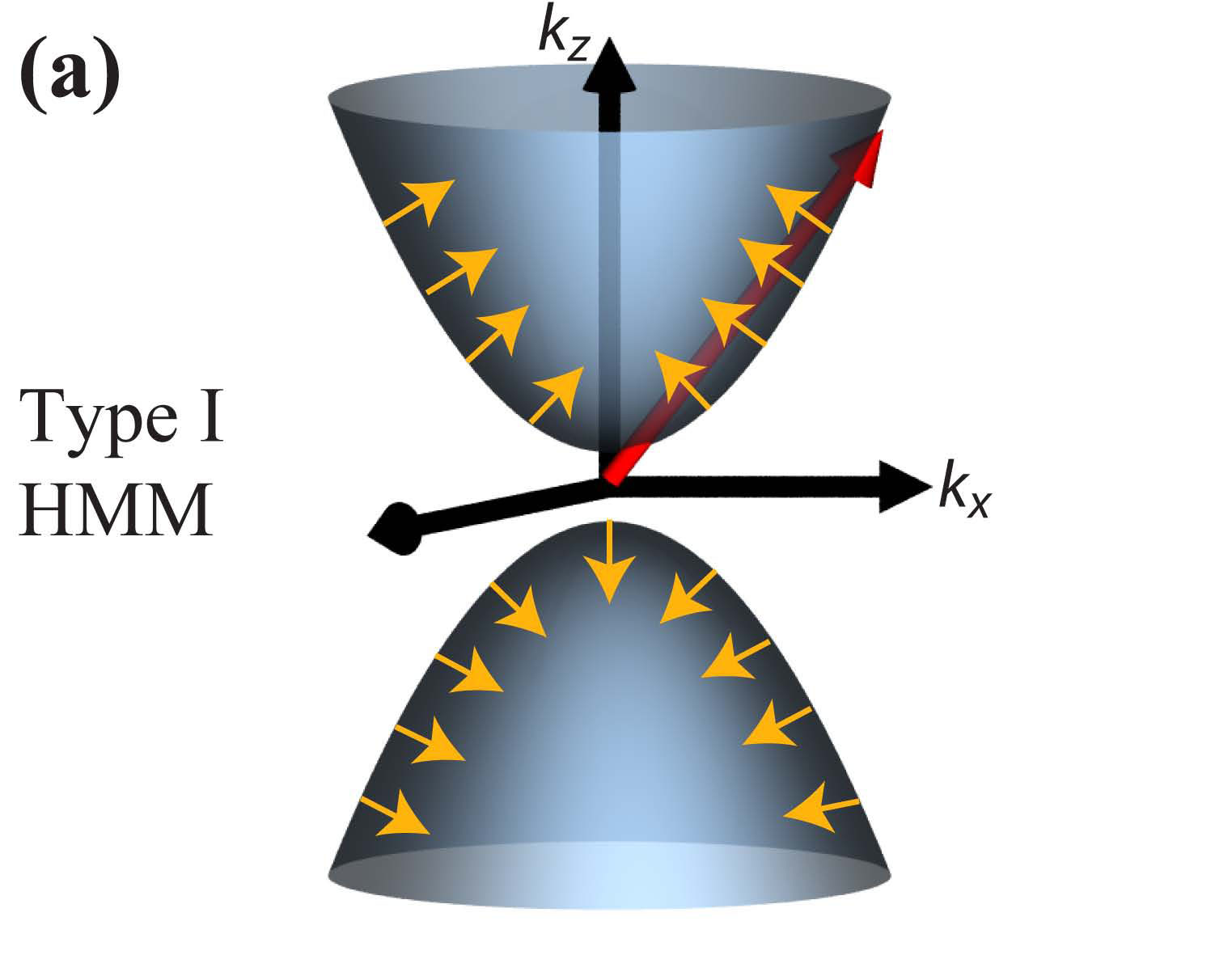}
\end{center}
\end{minipage}
\begin{minipage}[b]{0.15\linewidth}
\centering
\begin{center}
\includegraphics[width=30mm]{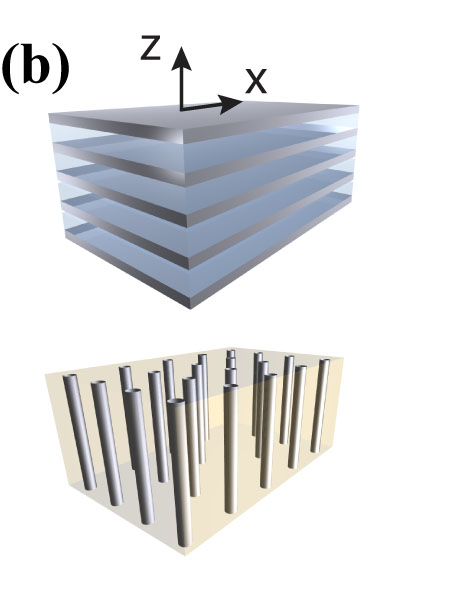}
\end{center}
\end{minipage}
\begin{minipage}[b]{0.275\linewidth}
\centering
\includegraphics[width=50mm]{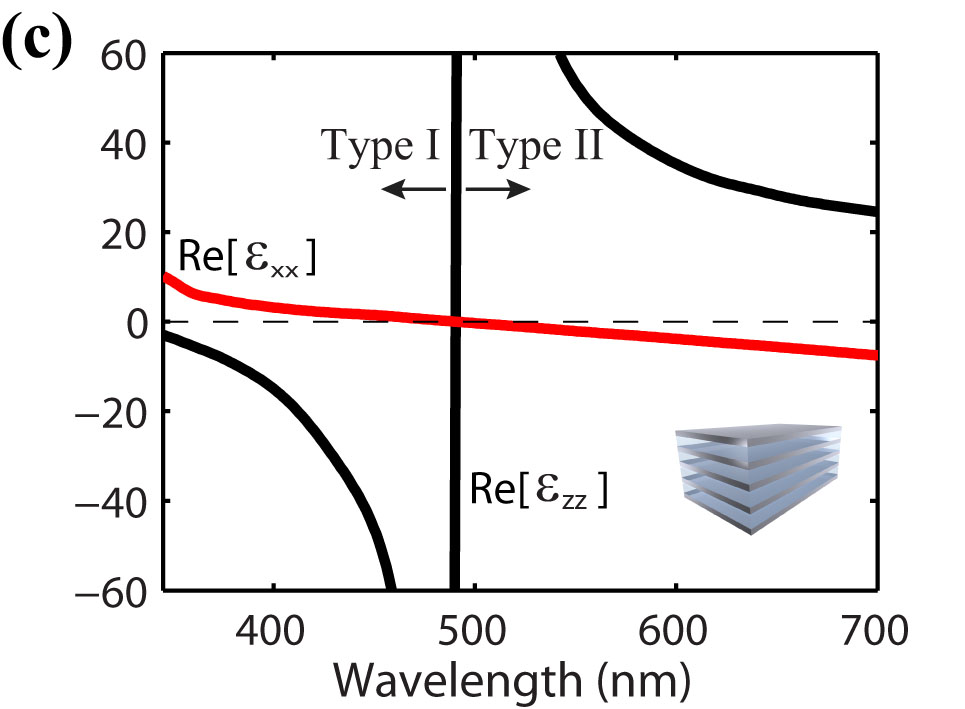}
\end{minipage}
\begin{minipage}[b]{0.25\linewidth}
\centering
\includegraphics[width=50mm]{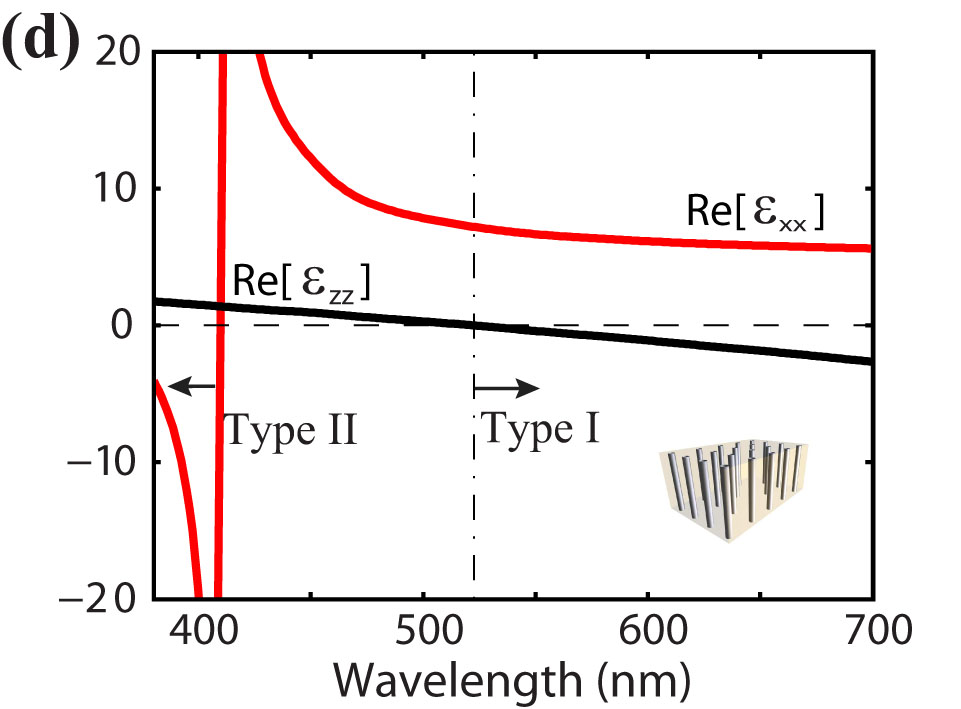}
\end{minipage}
\caption{(a) Isofrequency surface of type I HMM ($\epsilon_{xx} > 0, \epsilon_{zz} < 0$) with Poynting vectors (orange arrows) shown normal to the surface. Note that the Poynting vector component $S_x$ is negative and points in a direction opposite to the wavevector component. (b) The HMM response is achieved by using a multilayer realization of alternating subwavelength layers of metal and dielectric (shown above), or by using metal nanowires embedded in a dielectric host (shown below). (c) Real part of perpendicular and parallel components of dielectric permittivity predicted by EMT. Type I response occurs below $491$nm, while the Type II response ($\epsilon_{xx} < 0, \epsilon_{zz} > 0$) occurs in the longer wavelength region. We study a Ag/TiO2 multilayer structure with unit cell size $a$ = 30 nm (15nm/15nm).(d) Real part of perpendicular and parallel components of dielectric permittivity predicted by EMT for a nanowire system. Type I response occurs for a broader range of wavelengths above $522$nm, while the Type II response ($\epsilon_{xx} < 0, \epsilon_{zz} > 0$) occurs in the short wavelength region. The system consists of silver nanowires embedded in an alumina matrix with $\rho=0.22$ fill-fraction. 
\label{figure1} }
\end{figure*}

\section{Hyperbolic Metamaterials}

Our approach uses a non-magnetic uniaxial metamaterial, with dielectric tensor $\overline{\overline{\epsilon}}=$ diag$[\epsilon_{xx},\epsilon_{xx},\epsilon_{zz}]$, which exhibits hyperbolic dispersion for the extraordinary waves that pass through it. These waves are governed by the dispersion relation
\begin{equation}
(k_x^2 + k_y^2)/\epsilon_{zz} + k_z^2/\epsilon_{xx} = \omega^2/c^2
\label{dispersion_relation}
\end{equation}
where $k_x,k_y,k_z$ are the wavevector components of light inside the medium. The above equation produces a two-sheeted hyperboloid surface when $\epsilon_{xx} > 0$ and  $\epsilon_{zz} < 0$, referred to as a type I HMM. When $\epsilon_{xx} < 0$ and $\epsilon_{zz} > 0$, a single-sheeted hyperboloid is produced referred to as a type II HMM\cite{cortes_quantum_2012}. 

There are two main methods for achieving hyperbolic dispersion consisting of alternating metal-dielectric multilayers\cite{hoffman2007negative} or metal nanowires\cite{yao2008optical} in a dielectric host (see Fig. \ref{figure1}(b)). Effective medium theory (EMT) predicts these designs to achieve the desired extreme anisotropy in a broadband range provided the unit cells are significantly subwavelength (see Fig.\ref{figure1} (c) \& (d)). For a planar multilayer structure, EMT predicts the parallel and perpendicular permittivity components to be
\begin{equation}
\epsilon_{xx} = \rho \epsilon_m + (1-\rho)\epsilon_d 
\end{equation}
and
\begin{equation}
\epsilon_{zz} = \frac{\epsilon_m\epsilon_d}{ \rho \epsilon_d + (1-\rho)\epsilon_m},
\end{equation}
where $\epsilon_m$ is the metal permittivity, $\epsilon_d$ is the dielectric permittivity, and $\rho$ is the fill-fraction of the metal in the unit cell. The corresponding EMT parameters for a nanowire structure are
\begin{equation}
\epsilon_{zz} = \rho \epsilon_m + (1-\rho)\epsilon_d 
\end{equation}
and
\begin{equation}
\epsilon_{xx} = \frac{(1+\rho)\epsilon_m\epsilon_d+(1-\rho)\epsilon_d^2}{(1-\rho) \epsilon_m + (1+\rho)\epsilon_d}.
\end{equation}
We expect both the designs to support slow light modes, however we focus here on the multilayer design. The origin of slow light as well as hyperbolic dispersion are the coupled interface plasmon polaritons (or interface phonon polaritons) at the boundaries of the metal and dielectric layers \cite{elser2007nonlocal}. Note that the multilayer design does not support localized plasmons. This is especially important for active metamaterials where  amplification and stimulated emission of localized surface plasmons forms a major alternative competing channel to delocalized metamaterial modes\cite{stockman2011spaser}.

We note that previous studies have focused on a negative index metamaterial realization of slow light and associated effects\cite{tsakmakidis_trapped_2007,wu2011broadband}. In particular, Yao et. al.\cite{yao_ultrahigh_2009} have an excellent study on the topic of negative index metamaterial waveguides and enhanced Purcell Factors as well as Lamb Shifts. However, we note major differences of our study that render our realization conducive to experimental verification. i) Bulk optical magnetism which is required for double-negative metamaterials ($\epsilon < 0$ and $\mu < 0$) is a severe challenge leading to low figures of merit and narrowband operation. The multilayer and nanowire designs presented here does not use optical magnetism and only utilizes routine fabrication approaches leading to metamaterials with significantly better performance. ii) A more significant limitation of negative index approaches is near-field non-locality. Any metamaterial response is obtained by homogenization over a length scale larger than the unit cell size.  The large unit cell size of NIMs makes spatial dispersion a highly detrimental effect particularly for near-field effects such as LDOS enhancement\cite{yao_ultrahigh_2009}.  Thus even state of the art NIMs with unit cell sizes of 250 nm will show poor performance for any near field studies that require proximity less than 250 nm. iii) Artificial media that require $\epsilon < 0$ and $\mu < 0$ as considered in previous works are resonant metamaterials where effective medium theory cannot work in a broad bandwidth. The existence of unique modes such as slow light has to be ascertained through calculations that take into account the unit cell structure and not only effective medium approximations. Our non-magnetic ($\mu = 1$) design takes into account not only absorption and dispersion but also the finite patterning scale.  We include the role of near-field non locality  and obtain excellent  agreement with effective medium theory, a critical step towards experiment.

We emphasize that the HMM is highly tuneable for a wide range of wavelength regions by using different materials and layer thicknesses. Thus slow light in HMM waveguides presents a suitable platform for near-field studies and light matter interaction phenomena. Significant gain enhancement proportional to the enhanced density of states is expected at the slow light mode, making both the multilayer and nanowire design suitable for Van Hove singularity based metamaterial nano-lasers.

\section{Van Hove Singularity}
From the $k$-space topology in Fig.~\ref{figure1}(a), it is clear that the surface is unbounded. This results in arbitrarily large spatial frequency wavevector (high-$k$) modes and a density of states (DOS) that diverges in the effective medium limit\cite{jacob_broadband_2012}. These two features, along with the optical topological transition\cite{krishnamoorthy2012topological} that represents a sharp transition in the DOS, have been recently studied for device applications in nanophotonics\cite{cortes_quantum_2012,guo2012broadband,jacob2010engineering}. However, experimental demonstrations have only shown a modest increase in the LDOS by a factor of 2-5. 

Our focus in this paper is on a different mechanism for an order of magnitude larger enhancement in the local density of states. The density of states can be written as $\rho(\omega)=\int{\frac{dS}{|\nabla\omega(k)|}}$ in the low loss limit. Note that in electronic and photonic crystals, the points of high symmetry in k-space lead to a vanishing group velocity and corresponding Van Hove singularities in the density of states. This leads to observable consequences in  transport properties and  heat conduction in one dimensional electronic systems such as carbon nanotubes\cite{kim1999electronic}. A similar mechanism is related to band edge phenomena in photonic crystals\cite{dowling1994photonic}. In the low loss limit, the photonic density of states can be related to the inverse group velocity and hence slow light waveguide modes lead to features similar to Van Hove Singularities.  

We consider a multilayer system of silver ($Ag$) and titanium dioxide ($TiO_2$).  The layer thicknesses are 15nm each, with unit cell size $a=30$nm which is significantly subwavelength at optical frequencies but easily realizable experimentally. The waveguide slab thickness $d=$ 240nm consisting of 8 unit cells. EMT predicts that this structure will exhibit type I hyperbolic dispersion below 490.5nm (see Fig.~\ref{figure1}(c), only real parts have been shown). 

We define the LDOS in the near field normalized to its free-space value using a Green's function formalism  \cite{novotny2006principles}
\begin{equation}
\rho(\mathbf{r_o}, \omega_o) = \frac{2\omega_o}{\pi c^2}\mathrm{Im}\{Tr[\overline{\overline{G}} (\mathbf{r_o}, \mathbf{r_o}; \omega_o)]  \}.
\end{equation}
where $\mathbf{r_o}=(0,0,z_o)$ represents the position of the dipole located above the HMM waveguide, $\omega_o=2\pi c/\lambda$ is the free-space frequency and $c$ is the speed of light. The Green's tensor $\overline{\overline{G}}$ is related to the radiation field produced by an oscillating electric dipole source.
\begin{figure}[t!]  
\begin{center}
\includegraphics[width=70mm]{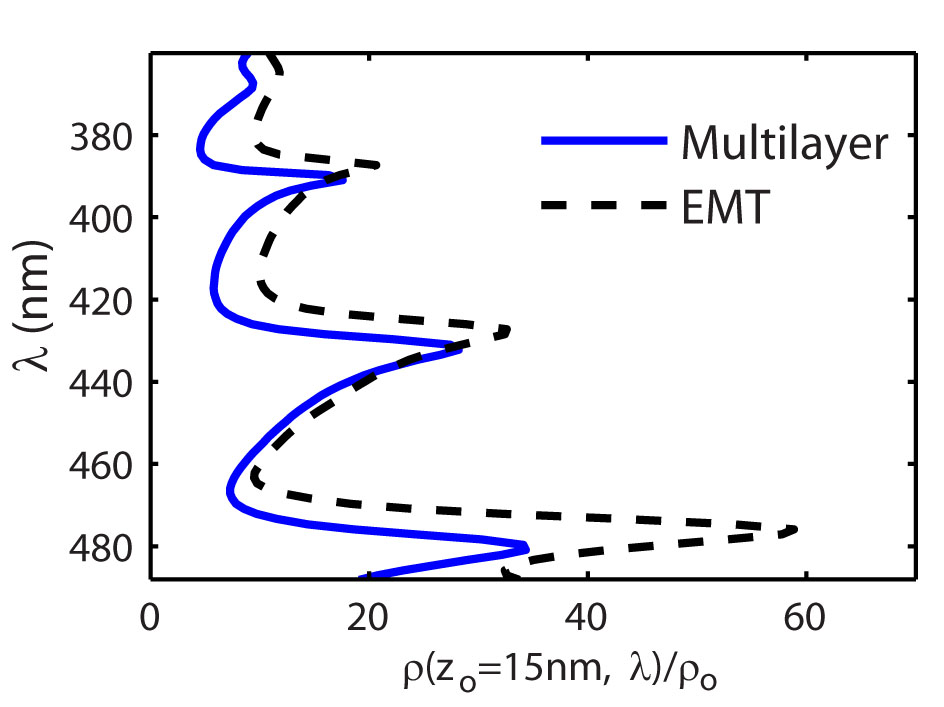}
\end{center}
\vspace{-10px}
\caption{Local density of states (normalized to free-space) at a distance \emph{$z_o=15nm$} away from HMM slab predicted by EMT. An excellent agreement seen for the multilayer structure calculated via transfer matrix method. We show the type I HMM wavelength region only which has not been studied before. Notice how the LDOS peaks occur at discrete wavelengths indicating that they must be fundamentally different from the usual broadband high-k modes in HMMs\cite{cortes_quantum_2012}.
\label{figure2A} }
\end{figure}

Fig.~\ref{figure2A} shows the LDOS enhancement in the type I HMM wavelength region at a distance of  $z_o=15$nm away from the HMM slab. Note the excellent agreement between the calculated LDOS for the homogenized metamaterial slab and the practical multilayer design. A number of recent experiments have focused on hyperbolic metamaterials and broadband density of states enhancement by a factor of 2-3. However, these experiments dealt with HMMs in the type II regime\cite{jacob2010engineering,krishnamoorthy2012topological} and the epsilon-near-zero regime\cite{noginov2010controlling}. We see two major differences for the aforementioned type I regime: i) an order of magnitude larger enhancements, and ii) discrete wavelengths at which the effect occurs. 

Also, note that the metamaterial LDOS peaks in Fig.~\ref{figure2A}  bear a striking resemblance to the Van Hove singularities that are often seen in electronic and photonic crystals. However, they are not a result of the periodicity of the structure, but rather they are predicted by the effective medium model of the metamaterial based on hyperbolic dispersion. In the next section, we will argue that these peaks can be physically attributed to slow light modes that lead to Van Hove singularities when $v_g\approx 0$ (low loss limit).

\section{Slow Light}
To understand the origin of these anomalous peaks, we define the wavevector-resolved local density of states (W-LDOS) as the LDOS distribution across the frequency spectrum and parallel wavevector ($k_x$) spectrum\cite{cortes_quantum_2012}, i.e.
\begin{equation}
\frac{\rho(\mathbf{r_o}, \omega_o)}{\rho_o} = \mathrm{Im}\int\limits_{0}^{~\infty} \frac{\rho(\mathbf{r_o},\omega_o,k_x)}{\rho_o} dk_x, 
\end{equation}
such that the W-LDOS $\rho(\mathbf{r_o},\omega_o,k_x)$ is given by
\begin{equation}
\frac{\rho(\mathbf{r_o},\omega_o,k_x)}{\rho_o}=  \frac{3}{2}\frac{i}{|\mathbf{p}|^2k_1^3} \frac{k_x}{k_z}\left\{ p_\perp^2 (1+r^p e^{i2k_z z_o})k_x^2 \right\}
\label{wavevector_ldos}
\end{equation}
for a perpendicular dipole with dipole moment $\mathbf{p}$ and fresnel reflection coefficient $r_p$ for $p$-polarized light.  This definition is a result of the Green's tensor which can be expanded as a summation of plane waves by using the Weyl identity.  For simplicity, we use the notation $k_x$ throughout the text however we take into account the entire radial component $k_{\rho}$ for the plane waves emitted by the point dipole. 

\begin{figure}[b!]
\begin{minipage}[b]{0.5\textwidth}
\begin{center}
\includegraphics[width=60mm]{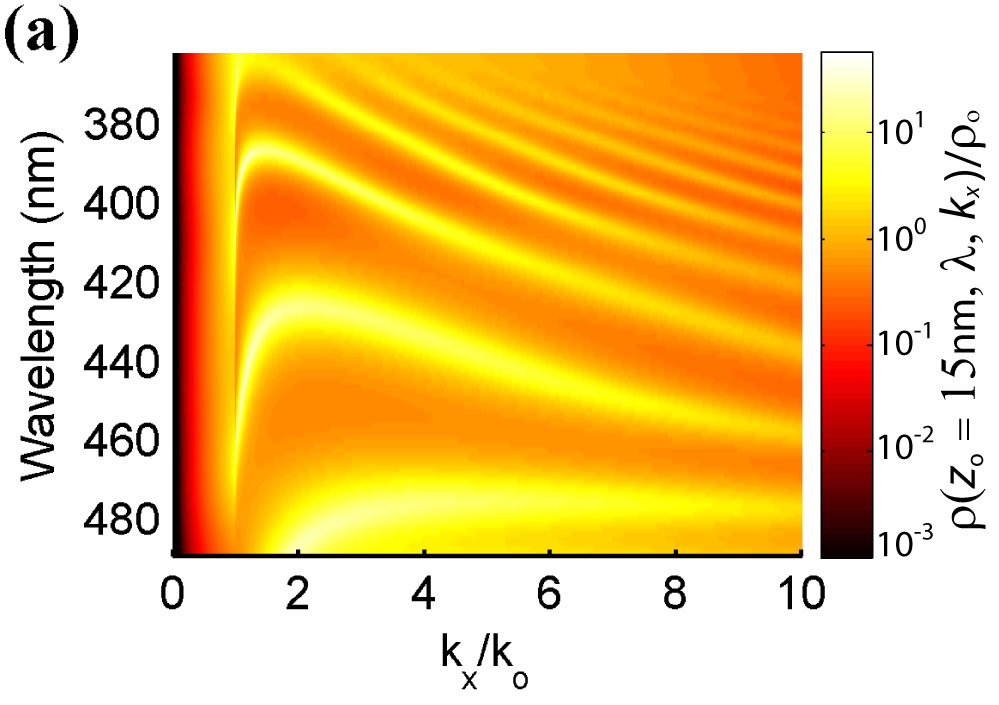}
\end{center}
\end{minipage}
\begin{minipage}[b]{0.5\textwidth}
\centering
\includegraphics[width=60mm]{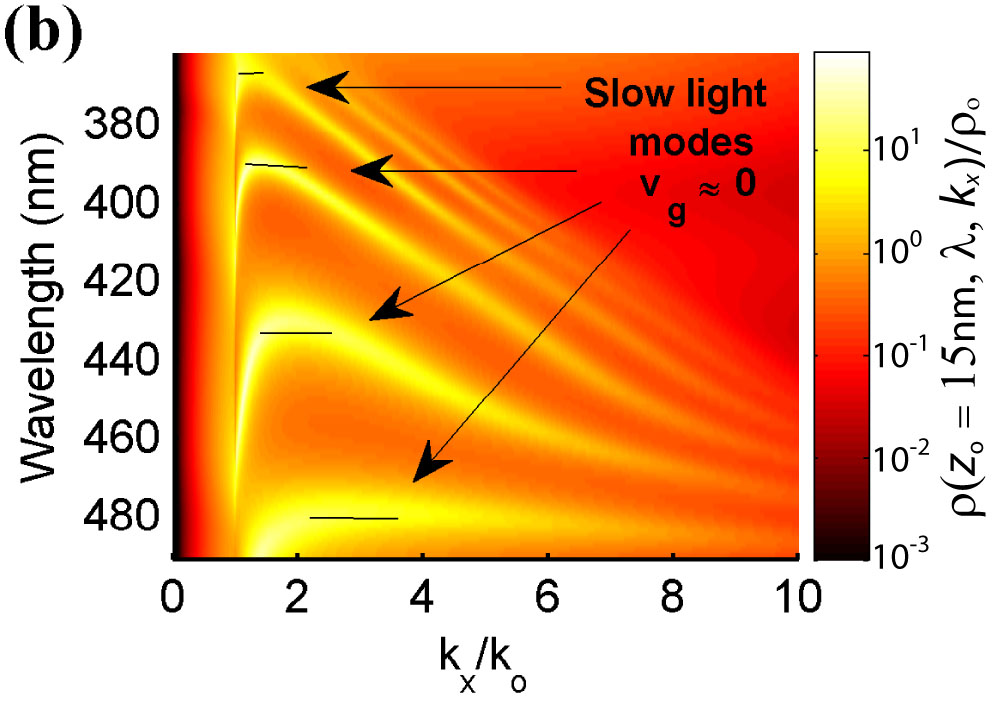}
\end{minipage}
\caption{Wavevector-resolved local density of states (normalized to free-space) of an emitter $15$nm away from HMM predicted by (a) effective medium theory. (b) Excellent agreement is seen for the practical multilayer structure. The highest LDOS occurs when the group velocity($\partial\omega/\partial k$), equal to the slope of the yellow regions, is nearly zero. These bright regions are the slow light modes that supported by the type I HMM.
\label{figure2}}
\end{figure}

Fig.~\ref{figure2} shows the log scale W-LDOS, normalized to free-space, for the homogenized effective medium as well as the practical multilayer system ($z_o=15$nm). Note that the bright bands in Fig.~\ref{figure2}(a) correspond to metamaterial modes which are bulk propagating waves in the effective medium limit. An excellent correspondence is seen with Fig.~\ref{figure2}(b) where these modes arise due to coupled plasmonic Bloch modes in the multilayer structure\cite{elser2007nonlocal}. At lower wavelengths and large wavevectors there is discrepancy between the two plots since effective medium theory is no longer valid for such modes\cite{cortes_quantum_2012}.

The brighest regions in Fig.~\ref{figure2} occur when the slope of the W-LDOS bands is equal to zero. The density of states is proportional to the inverse of the group velocity of light, $\rho(\lambda) \propto \partial k/\partial \omega \propto 1/v_g$ , therefore we attribute the LDOS peaks in Fig.~\ref{figure1}(c) to slow-light modes with a group velocity nearly equal to zero. This is in accordance with the work of Yao.et.al\cite{yao_ultrahigh_2009}, which predicted LDOS peaks for a ($\epsilon<0$ and $\mu < 0$) metamaterial as a result of slow light modes as well. As we mentioned earlier, however, optical magnetism is quite difficult to achieve in practive, thus we provide an alternative approach based on metamaterials with hyperbolic dispersion. Also, notice that in the ideal case of $v_g=0$,  these modes would correspond to Van Hove singularities in the PDOS at discrete wavelengths\cite{hess_active_2012} which are fundamentally different from the broadband singularity in the density of states that is a result of the hyperbolic dispersion\cite{jacob_broadband_2012}. 


We now turn to the physical nature of the slow light mode supported by the HMM. In the low absorption limit, the Poynting vector is normal to the isofrequency surface (see Fig.~\ref{figure1}(a)). From this picture it is clear that the component of the Poynting vector along the $x$-direction ($S_x$) can be of opposite sign to the wavevector component ($k_x$) such that $k_x\cdot S_x <0$.   This argument also remains true with absorption because the Poynting vector, $\vec{S} \propto k_x/\epsilon_{zz}\hat{x} + k_z/\epsilon_{xx}\hat{z}$, satisfies the relation $Re(k_x)\cdot Re(S_x) <0$ \cite{smith2004negative}. Propagating modes in the waveguide must have the same parallel wavevector ($k_x$) component in all three mediums in order to satisfy the boundary conditions. We define $Re(k_x)$ to be positive which consequently makes the energy flow inside the HMM negative. As a result, it is clear that the slow light waveguide mode is due to a negative energy flow ($k_x\cdot S_x <0$) inside the HMM which negates the positive energy flow ($k_x\cdot S_x >0$)  of the outer  dielectric claddings \cite{alekseyev_slow_2006}. This directly contrasts other slow-light effects seen in PhCs or EIT systems that rely on a standing wave condition or an atomic resonance of the material respectively\cite{baba2008slow,khurgin_slow_2010}.


\section{Complex Band Structure}

Next, we analyze the effect of losses and dispersion on the slow-light modes by analyzing the complex-$k_x$ dispersion relation of the hyperbolic metamaterial waveguide. The complex-$k_x$ picture shows the detrimental effect of loss and is relevant for applications that utilize a continuous pump source to excite the modes of the system. Note that while the W-LDOS plots in Fig.~\ref{figure2} do take loss into account, the resolution is performed with respect to the real wavevector $k_x$. 

The waveguide modes inside an HMM are described in general by the transcedental equation: 
\begin{equation}
tan(k_{z,2}d) = \frac{\epsilon_{xx} k_{z,2}(\epsilon_3 k_{z,1}+\epsilon_1 k_{z,3})}{(\epsilon_3\epsilon_1 k_{z,2}^2 - \epsilon_{xx}^2 k_{z,1}k_{z,3})}, 
\end{equation}
where $k_{z,i}$ is the perpendicular wavevector component inside medium $\epsilon_i$ ($i$=1,2,3)\cite{alekseyev_slow_2006}. Medium 1 ($\epsilon_1=1$) is the half-space of the emitter, medium 2 is the HMM slab, and medium 3 ($\epsilon_3=1$) is the substrate half-space.

\begin{figure}[t!]
\begin{minipage}[b]{0.50\textwidth}
\begin{center}
\includegraphics[width=65mm]{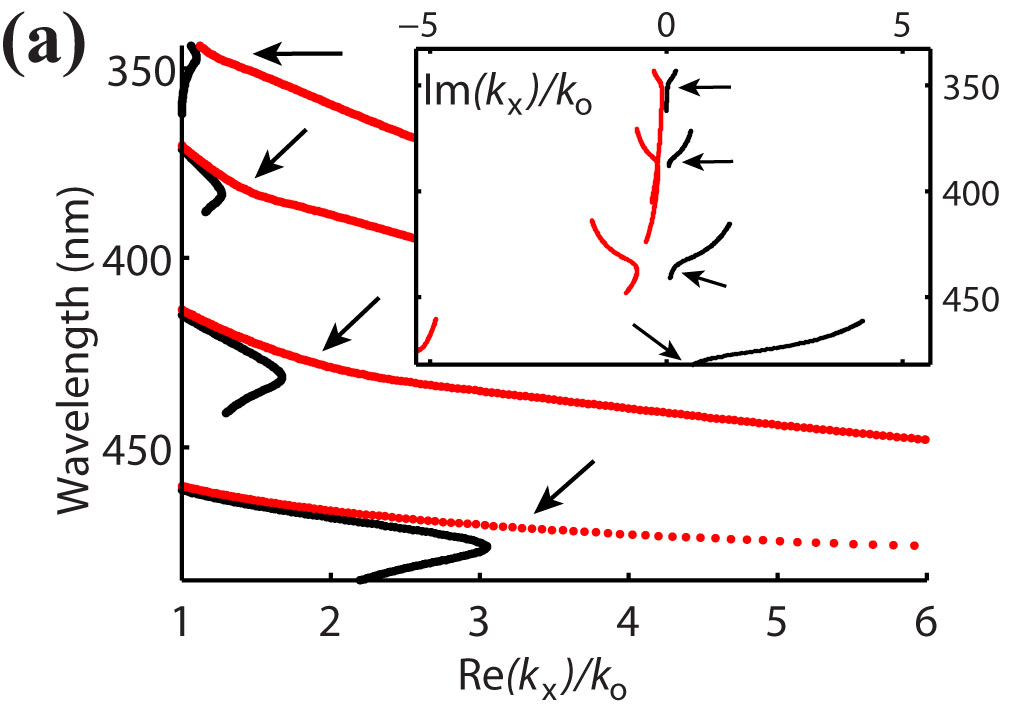}
\end{center}
\end{minipage}
\begin{minipage}[b]{0.5\textwidth}
\begin{center}
\includegraphics[width=65mm]{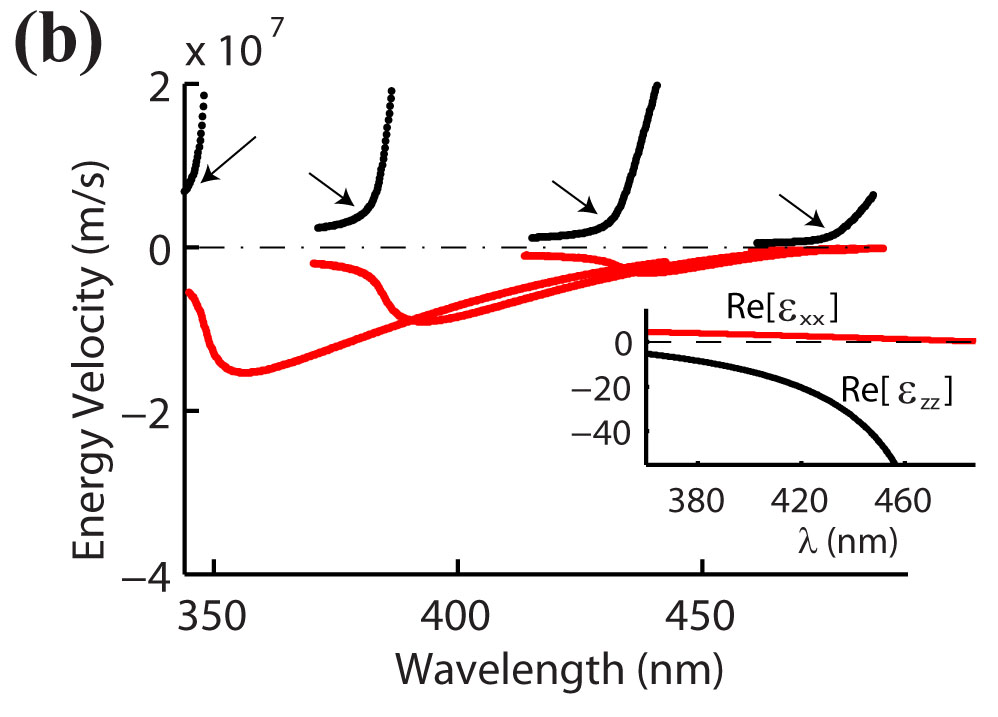}
\end{center}
\end{minipage}
\vspace{-4mm}
\caption{Complex-$k_x$ dispersion relation of slow-light modes predicted by EMT, calculated numerically (a) $\lambda$ vs. $Re(k_x)$  and $\lambda$ vs. $Im(k_x)$ (shown in inset). Note that the waveguide mode solutions split into two branches: a forward travelling wave (black branch) and backward travelling wave (red branch). A negative real part of the wavenumber has to be chosen for this backward branch so that $Im(k_x) > 0$ (b) Energy velocity is calculated numerically using values from dispersion relation. The arrows point towards slow-light mode region before it becomes a highly leaky slow light mode.
\label{figure4A}}
\end{figure}

\begin{figure*}[t!]
\raggedright
\begin{minipage}[b]{0.3\linewidth}
\begin{center}
\includegraphics[width=49mm]{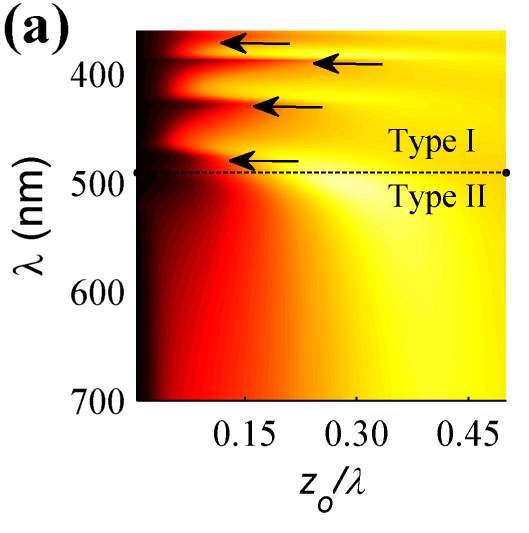}
\end{center}
\end{minipage}
\begin{minipage}[b]{0.3\linewidth}
\centering
\includegraphics[width=56mm]{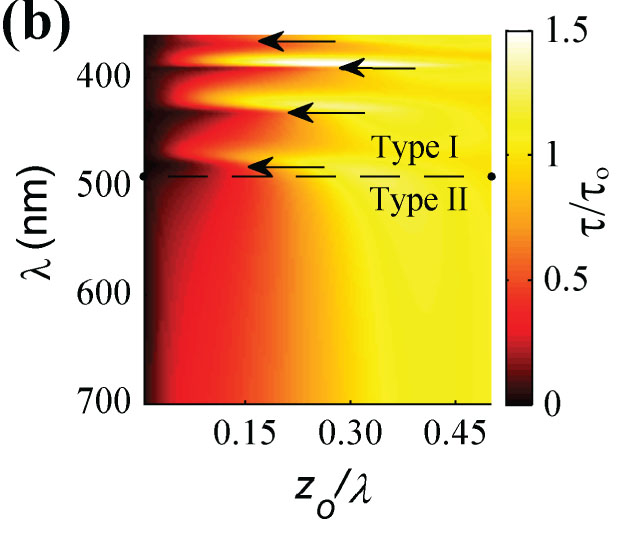}
\end{minipage}
\hspace{0.2cm}
\begin{minipage}[b]{0.3\linewidth}
\centering
\begin{center}
\includegraphics[width=66mm]{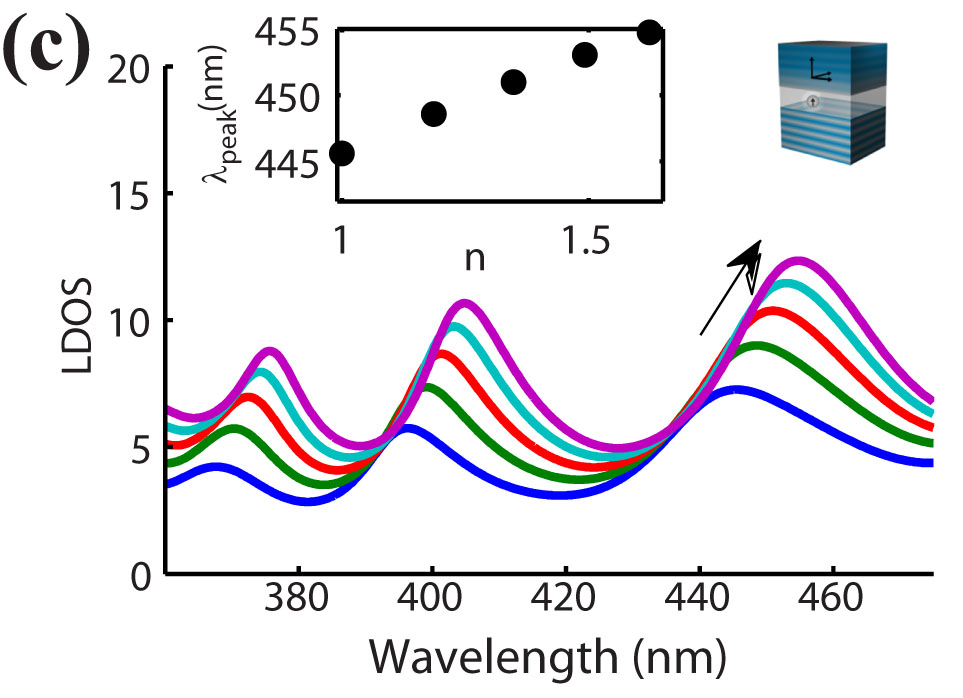}
\end{center}
\end{minipage}
\vspace{-1mm}
\caption{Near field variation of the spontaneous emission lifetime of quantum emitters near the slow light HMM waveguide. The distance of the quantum emitter is varied and normalized to the operating wavelength. (a) Numerical calculation of EMT and (b) practical multilayer system. The dark regions (colorbar $< 0.2$) corresponding to lower lifetime  of the slow-light modes extend much further denoting excellent coupling of the emitters to slow light modes as opposed to other radiative and non-radiative channels. (c) The refractive index of an $80$nm embedded layer between two $240$nm HMM slabs (shown in inset) is varied to demonstrate sensitivity of slow-light mode, useful for optical biosensing. Inset shows the linear relation of the $450$nm slow-light mode as a function of the refractive index \emph{n} equal to: 1.0, 1.2, 1.3, 1.5 and 1.6 (follows black arrow in main plot).
\label{figure5}
}
\end{figure*}

The solution to the HMM waveguide equation is shown in Fig.~\ref{figure4A}(a). Note that Fig.~\ref{figure4A}(a) follows very closely with the yellow regions of Fig.~\ref{figure2}. However, the key difference lies near the slow light mode wavelengths (denoted by the black arrows) where it is clear from this plot that the modes split into two branches: a forward travelling wave (black branch) where the Poynting vector and wavevector point in the same direction $k_x\cdot S_x >0$, and a backward travelling wave (red branch) where $k_x\cdot S_x <0$. A negative real part of the wavenumber has to be chosen for this backward branch so that $Im(k_x) > 0$. 
These two branches get closer together near the slow-light mode wavelength (denoted by the black arrows) and form a leaky slow-light wave that eventually merges to the bulk plasma oscillation branch of the metal at ($\omega \approx \omega_p$ and $k_x\approx0$). Note that the leaky modes are very lossy and have short propagation length as seen by the $Im(k_x)$ plot (inset Fig.~\ref{figure4A}(a)). This interesting behavior of slow light modes has been noted  in media with $\epsilon < 0$ and $\mu < 0$\cite{yao_ultrahigh_2009}. 

In an absorbing, dispersive medium the group velocity loses its true physical meaning and  becomes ill-defined\cite{reza_can_2008}. This has been an important issue while undertanding slow-light modes and so we use energy velocity $v_E$ as the correct physical quantity\cite{fedyanin_backward_2010, fedyanin_stored_2010}. $v_E$ is defined as the ratio of the time-averaged Poynting vector $S_x$ and electromagnetic energy density $u_{em}$, where both are integrated along the $z$-direction to obtain the contribution along the propagation direction of the waveguide mode
\begin{equation}
 v_E = \frac{\int^\infty_{-\infty} Re(S_x) dz}{\int^\infty_{-\infty} u_{em} dz}.
 \label{v_energy}
\end{equation}
For the calculation of Eqn. (10), we first found the analytical expressions for the time-averaged Poynting vectors inside the HMM and outer media. Then, we used a well-defined (positive definite) electromagnetic energy density for HMMs taking into account absorption and dispersion\cite{oughstun_velocity_1988,nunes_analysis_2012}. In order to use the formalism used in Ref. 38, we require the parallel and perpendicular components of the permittivity to be defined by either a Drude or Lorentz model. For an HMM based on a planar multilayer structure, $\epsilon_{xx}$ can be modeled by a Drude model with plasma frequency $\omega_P$, while $\epsilon_{zz}$ can be modeled by a Lorentz model
\begin{equation}
\epsilon_{xx}= \epsilon_{o1} - \frac{\omega_P^2}{\omega^2 + i\Gamma\omega}
\end{equation}
\begin{equation}
\epsilon_{zz}= \epsilon_{o2} + \frac{\omega_P^2}{\omega_1^2 - \omega^2 - i\Gamma\omega},
\end{equation}
where $\epsilon_{o1}=9.7$, $\epsilon_{o2}=6$, $\Gamma=6.5\times10^{13} s^{-1}$, $\omega_1=3.84\times10^{15} s^{-1}$ and $\omega_P=1.19\times10^{16} s^{-1}$.
We obtained values for $\epsilon_{xx}$ and $\epsilon_{zz}$ based on a fitting to the EMT result in Fig.~\ref{figure1}(c). This definition was used to calculate the results in Fig.~\ref{figure4A}(a) and (b).

Fig.~\ref{figure4A}(b) shows the energy velocity for all four slow-light modes as a function of wavelength. The waveguide modes slow down at specific wavelengths (denoted by the black arrows) before becoming leaky modes, showing that stopped-light is not possible in the complex-$k_x$ case. Our result for hyperbolic metamaterials is in agreement with previous results of $\epsilon < 0$, $\mu < 0$ media\cite{yao_ultrahigh_2009}. Light reaches an average minimum velocity of $c/100$ around the slow-light region while also reaching the global minimum of $c/1000$ in the region of highest dispersion, $\lambda \approx 480$nm. It is important to note that for time-dependent processes where modes are excited by an optical pulse, the complex-$k_x$ picture is not sufficient. Nevertheless, we have confirmed that solving the complex-$\omega$ dispersion relation gives similar results.\cite{yao_ultrahigh_2009,tsakmakidis_trapped_2007}

\section{Applications}

We now discuss multiple applications of Van Hove singularities in multilayer and nanowire hyperbolic metamaterials. 

\subsection{Lifetime Engineering}
One approach to observing the strong LDOS enhancement effect and unique spectral signatures of the Van Hove singularities is through the spontaneous emission lifetime of emitters such as quantum dots and dye molecules. The scaling dependence of the  lifetime to the distance ($z_o$) due to near-field coupling with the slow-light modes is markedly different\cite{yao_ultrahigh_2009} from other photonic channels like quenching, non-radiative decay, plasmon modes and conventional HMM modes\cite{ford_electromagnetic_1984}. 

We compare the scaling dependence between the slow-light modes in the type I region with the high-$k$ modes in the type II wavelength region (see Fig.~\ref{figure5} (a) and (b)). We see that the slow-light modes have lifetimes that remain quite small, denoted as the dark regions ($<0.2$ in the colorbar), for much larger distances than the previously studied high-$k$ modes\cite{jacob_broadband_2012}. This is due to the fact that slow light modes are much closer to the light line ($k_x/k_o=1$) than the high-$k$ modes and so they do not decay as quickly outside the HMM. This marked difference from previous experiments\cite{jacob2010engineering,krishnamoorthy2012topological} should be discernible in near-field lifetime studies.

Furthermore, we have calculated that approximately $90\%$ of the total contribution of the LDOS peaks (in Fig.~\ref{figure2A}) is due to the slow-light modes of the type I HMM, i.e $\tau_{_{SL}}/\tau \approx 0.9$. However, these modes are lossy and hence would be eventually absorbed in the metamaterial. Only 10\% of the light is reflected from the metamaterial and reaches the far-field. The figure of merit for the slow-light mode, considering realistic losses and dispersion, is $Re(k_x^{sl})/Im(k_x^{sl}) \approx 15$ where $k_x^{sl}$ is the wavector of the slow light mode. This demonstrates that the majority of the enhancement factors are a direct result of the metamaterial VHS, and not the high-$k$ modes of the HMM. The VHS contribution could be measured by a far-field detector by using appropriate out-coupling techniques to convert the confined slow-light mode into propagating waves. 


\subsection{Optical Sensing}
Another potential application is shown in Fig.~\ref{figure5} (c), which shows the sensitivity of slow light modes to refractive index changes -- an important feature for the optical sensing of biological molecules or compact devices for gas monitoring\cite{jensen2008slow} . We modeled the LDOS between two 240nm EMT HMM slabs, with an embedded layer thickness of 80nm (see Fig.~\ref{figure5}(c) inset). This can be probed for example by  embedded fluorescent molecules or absorbed gases.  

Fig.~\ref{figure5}(c) shows the LDOS enhancement normalized to free-space for different refractive indices $n$ of the embedded layer. Note that the peaks red shift as the refractive index increases. The sensitivity, given by larger shifts, also increases in relation to the increasing dispersion in the longer wavelength range\cite{khurgin_slow_2010}. The inset shows a linear variation of 450nm wavelength peak within this refractive index regime. The linearity is valid in the first order approximation when the refractive index changes are not very large. The effective index of the slow light varies slowly with small perturbations of the homogenized metamaterial response. 

\subsection{Thermal Engineering}

We expect a major application of the predicted narrowband enhancement in the LDOS to be in near-field thermal emission beyond the black body limit. Recently, it has been shown that hyperbolic metamaterials can provide broadband super-planckian thermal emission\cite{guo2012broadband} that is useful for thermal management and coherent thermal sources. Thermal energy transfer using van hove singularities will be important for near-field thermophotovoltaics \cite{molesky2013high} where narrowband super-planckian emission matched to a photovoltaic cell is required. Note that near-field thermal emission spectroscopy can detect the unique spectral signatures of these Van Hove Singularities.  Our recent work shows that the metamaterial Van Hove singularity presented here leads to narrowband super-planckian thermal emission\cite{guo2013high} which can be detected by near-field thermal emission spectroscopy\cite{jones2012thermal}. A detailed analysis of near-field thermal engineering using Van Hove singularities will be published elsewhere.

\subsection{Active Devices}

A final important application of the metamaterial VHS will be in active plasmonic devices\cite{hess_active_2012,ni_loss-compensated_2011,boardman_active_2011}. In such devices, large gain becomes necessary to compensate for metallic losses that hinder the device performance of many plasmonic systems. The VHS is ideal for these applications because it provides a very large group index leading to large gain enhancement factors\cite{grgic2012fundamental}. Controlling the decay channels of the gain medium can minimize the threshold for nano-lasing \cite{ryu2002very}. Also, the large PDOS of the VHS provides a dominant channel for which the gain medium emit into, thus achieving optimal control of gain medium's emission.We emphasize that both multilayer and nanowire hyperbolic metamaterials can be used for active devices and van hove singularity nano-lasers.   

\section{Conclusion}
To conclude, we have introduced the photonic analogue of the electronic Van Hove singularity in metamaterials based on hyperbolic dispersion. We have shown that they exhibit a very high LDOS despite material absorption and dispersion. We also highlighted some important applications of the metamaterial VHS through lifetime and active devices. The VHS also showed high sensitivity to refractive index changes opening the possibility of nanowaveguide sensors. The multilayer platform that we have presented is accessible to many experimentalists, and is suitable for future studies focusing on the realization of stopped-light in active media. This work will also be important for other applications in non-linear metamaterials and quantum optics.


\section{Acknowledgments}
The authors would like thank A. Fedyanin for insightful discussions.

\bibliographystyle{unsrt}
\bibliography{SlowLight_V8}

\end{document}